\documentclass[10pt,notitlepage]{iopart}
\usepackage{a4,epsfig}
\usepackage{url}
\usepackage[dvipsnames]{xcolor}
\usepackage[hidelinks]{hyperref}
\usepackage{lineno}
\usepackage{multirow}
\usepackage{xcolor}
\usepackage[nosort]{cite}
\usepackage{iopams}
\usepackage[toc,page,title]{appendix}
\usepackage[normalem]{ulem}
\usepackage{graphicx}
\usepackage{subfig}

\graphicspath{{./pictures/}}

\let\orgautoref\autoref
\renewcommand{\autoref}
        {\def\subfigureautorefname{Figure}%
         \orgautoref}

\begin{document}

\title{The Virgo Newtonian calibration system for the O4 observing run}

\author{
F.~Aubin$^1$\footnote{Corresponding author: florian.aubin@iphc.cnrs.fr}, E. Dangelser$^1$, D.~Estevez$^1$, A.~Masserot$^2$, B.~Mours$^1$, T.~Pradier$^1$, A.~Syx$^1$, P.~Van Hove$^1$
}

\address{$^1$Université de Strasbourg, CNRS, IPHC UMR 7178, F-67000 Strasbourg, France}
\address{$^2$Univ. Savoie Mont Blanc, CNRS, Laboratoire d'Annecy de Physique des Particules - IN2P3, F-74000 Annecy, France}

\begin{abstract}
After initial tests performed during previous observing runs, a Newtonian Calibrator (NCal) system was developed and installed on the Virgo gravitational wave detector for the O4 observing run. This system, which is continuously operated, provides the absolute calibration of Virgo for this run. Its $1\mbox{-}\sigma$ uncertainty of $0.17\%$ on the amplitudes of the injected signals is better than that obtained with other calibration techniques like the photon calibrator (PCal). This paper presents this NCal system and details the different sources of uncertainties.
\end{abstract}
\newpage


\section{Introduction} \label{sec:introduction}
Since the first direct gravitational wave detection in September 2015 \cite{GW150914}, the rate of observed events increased with each new observing run, reaching few events per week with the ongoing fourth observing run of the Advanced LIGO \cite{AdvLIGO}, Advanced Virgo \cite{AdvVirgo}, and KAGRA \cite{KAGRA1, KAGRA2} network. 
This rate increase is expected to continue with the planned upgrades of the existing detectors, as well as the foreseen third generation detectors \cite{ET, CE}. 
This transition from a discovery mode to precision science puts strong constraints on the calibration of gravitational wave detectors to produce unbiased results when measuring the Hubble constant\cite{H0O2}, rates of astrophysical sources\cite{O3pop} or testing General Relativity\cite{TGRO3}. 
\\

The calibration of gravitational wave detectors is performed by injecting known signals, and measuring their effects on the output signal. 
To do so, it is necessary to build actuators that induce an accurate displacement of the test masses of the interferometers.

During the previous observing run, O3, the Virgo detector was calibrated using the so called Photon Calibrator (PCal) which induces a displacement of the end-of-arm mirrors using the radiation pressure of an auxiliary laser \cite{PCalO3}. 

After a first test at the end of the O2 run, a new prototype of Newtonian Calibrator (NCal) was tested during O3. This device creates a modulated gravitational field produced by rotating masses to induce a displacement of the mirrors of the Virgo detector.
A finite element analysis program, named FROMAGE, was used to predict the expected signal of the NCal actuators \cite{FROMAGE}. 
A precise characterization of the NCal setup is mandatory to minimize the systematic uncertainties. 
The accuracy of the O3 test NCal system was $1.4\%$, similar to the PCal system, with a difference of $3\%$ between the two techniques \cite{Virgo_Cali}.
A NCal system was also tested at LIGO Hanford during O3, reaching a total uncertainty of $0.8\%$ in a more limited bandwidth \cite{NCal_Ligo}.

It was decided to design and install an improved Virgo NCal system for the O4 run with the goal of providing the absolute calibration of the detector and with continuous operation. 
Since the NCal could only operate in a limited frequency band (up to $150~\mathrm{Hz}$), the PCal system remains mandatory to calibrate the full frequency band of Virgo (up to a several thousand hertz), once cross-calibrated with the NCal system.
PCal systems have also been improved, reaching for LIGO a combined standard uncertainty level as low as $0.3\%$ for the induced mirror displacement \cite{LIGOPCal}.
\\

This paper is structured to address all sources of uncertainty that may affect the signal injected by the Virgo O4 NCal system, with all uncertainty values provided at the $1\mbox{-}\sigma$ confidence interval.
Section \ref{sec:O4system} of this paper outlines the NCal setup installed on the Virgo detector for the O4 run. 
It comprises six rotors disposed on three axes around the north-end mirror.
The choice of such a layout was motivated by the need to minimize uncertainties arising from actuator positioning relative to the mirror location, as described in section \ref{sec:setupUncertainties}.
Section \ref{sec:rotorUncertainties} details how rotor geometry and material uncertainties are taken into account in signal prediction using the finite element simulation FROMAGE.
While the original actuators were made of aluminum, parasitic coupling studies presented in section \ref{sec:couplingUncertainties} revealed that this material tends to produce a magnetic field that couples to the magnets attached to the mirror. 
For this reason, the rotors nearest to the mirror were changed to PVC rotors during the commissioning phase.
Finally, in section \ref{sec:allUncertainties}, we provide an assessment of the overall error budget for the displacements induced by the NCal system, followed by a conclusion and a brief discussion on the future plans for the Virgo NCal system.

\section{The O4 NCal system} \label{sec:O4system}
\subsection{System description}

\begin{figure}[h!]
    \centerline{\includegraphics[width=1\textwidth]{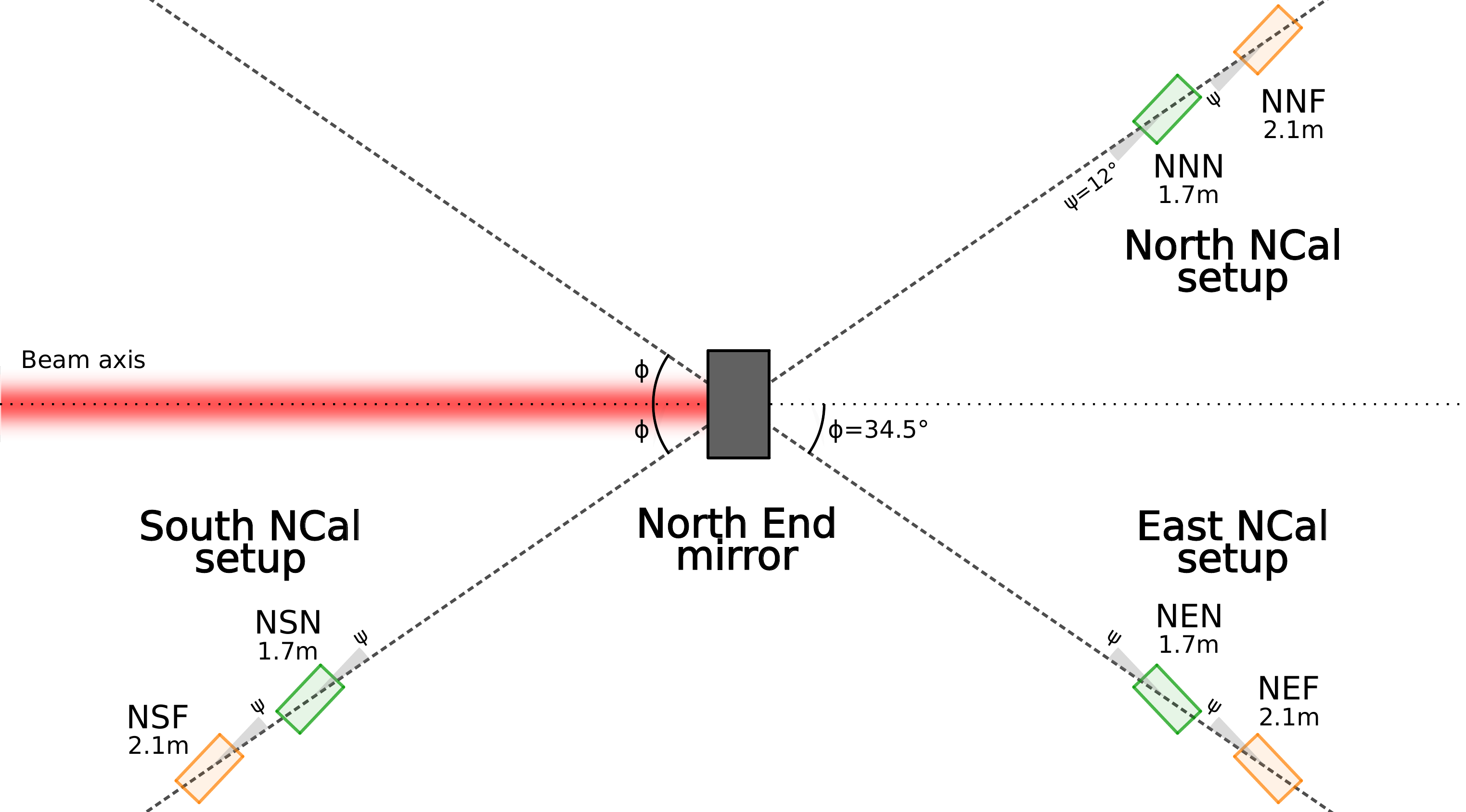}}
    \caption{Newtonian calibrator system of the Virgo interferometer at the start of O4b. 
    Each rotor is identified by three letters: the beam direction (N for North), the NCal position relative to the mirror (N for North, S for South and E for East), and the triplet identifier (N for Near and F for Far). 
    The green rotors (Near) are made of PVC, while the orange rotors (Far) are in aluminium.}
    \label{fig:O4setup}
\end{figure}

One of the most critical parameters of the NCal system is its position relative to the interferometer mirror.
However, since the mirror is inside a vacuum chamber, this parameter is difficult to measure and therefore poorly known. 
Previous work \cite{NCalO2, NCalO3} has shown that a pair of NCals located around the mirror can be used to infer this position, therefore significantly reducing the associated uncertainty.
Based on this result, we chose a Virgo NCal configuration for O4 that consists of two triplets of actuators positioned around the north-end mirror of the interferometer.
A top view of this system is shown in \autoref{fig:O4setup}.
Each triplet includes rotors installed on either side of the mirror, called North and South.
To fully measure the mirror location in the interferometer plane and perform additional measurements (see section \ref{sec:couplingUncertainties}), triplets also includes a third rotor, called East, installed symmetrically to the North one relative to the laser beam axis.
One triplet (in orange on figure \ref{fig:O4setup}) equipped with aluminium (Al 7075-T6) rotors, named Far, is positioned at $2.1~\mathrm{m}$ of the north-end Virgo's mirror. 
In order to limit the effects of the magnetic field produced by such rotors, discussed in \autoref{sec:couplingUncertainties}, the rotors of the second triplet named Near (in green on figure \ref{fig:O4setup}), at $1.7~\mathrm{m}$ of the mirror, are made of polyvinyl chloride (PVC).
These two triplets offer independent measurements and redundancy for the whole calibration system.
 
\begin{figure}[h]
    \centerline{\includegraphics[width=0.5\textwidth]{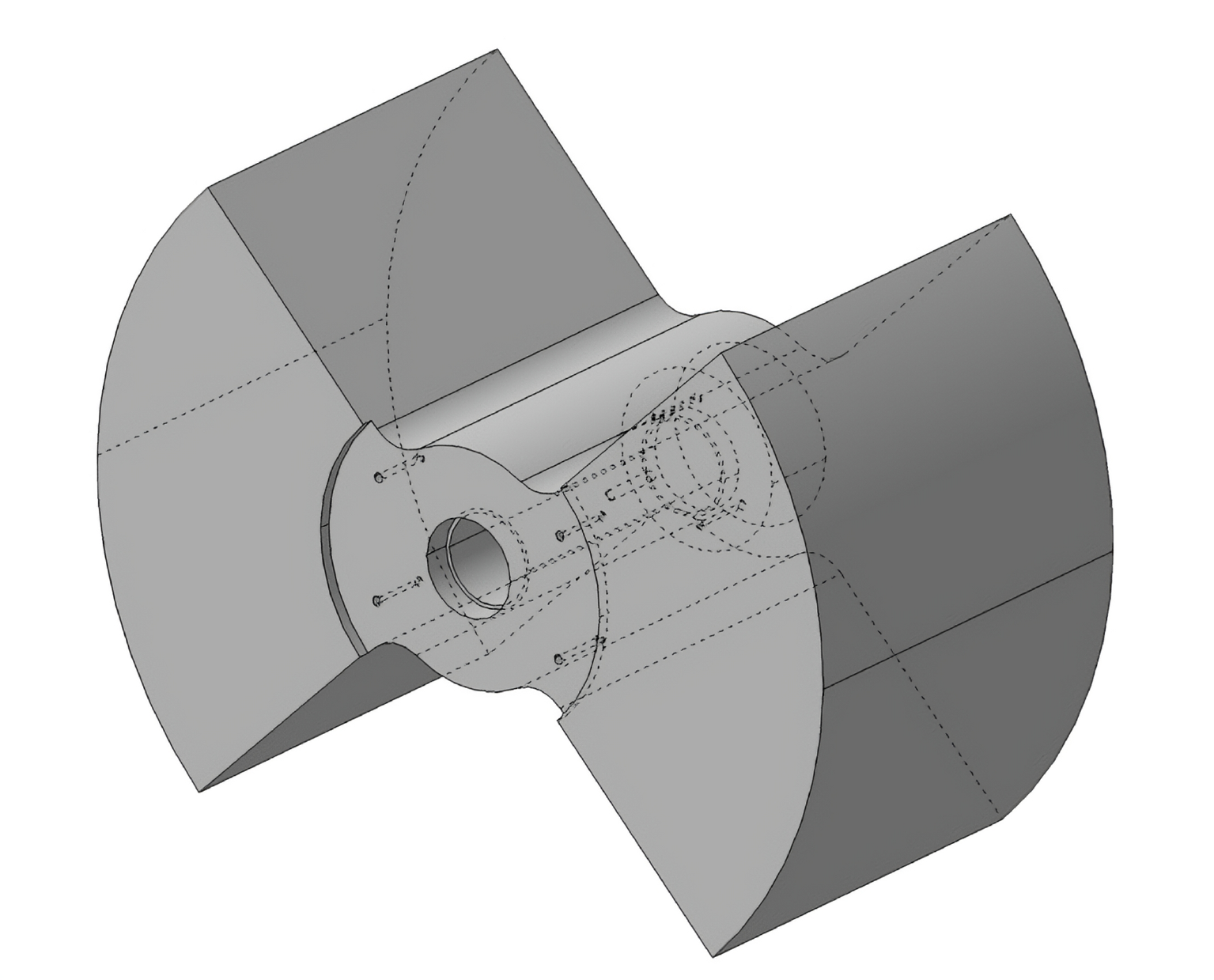}}
    \caption{Isometric view of a rotor. 
    The diameter is 208 mm. The rotation axis lies in the interferometer plane.}
    \label{fig:rotor}
\end{figure}

\begin{figure}[h!]
  \centering
  \subfloat[]{\includegraphics[width=0.5\textwidth]{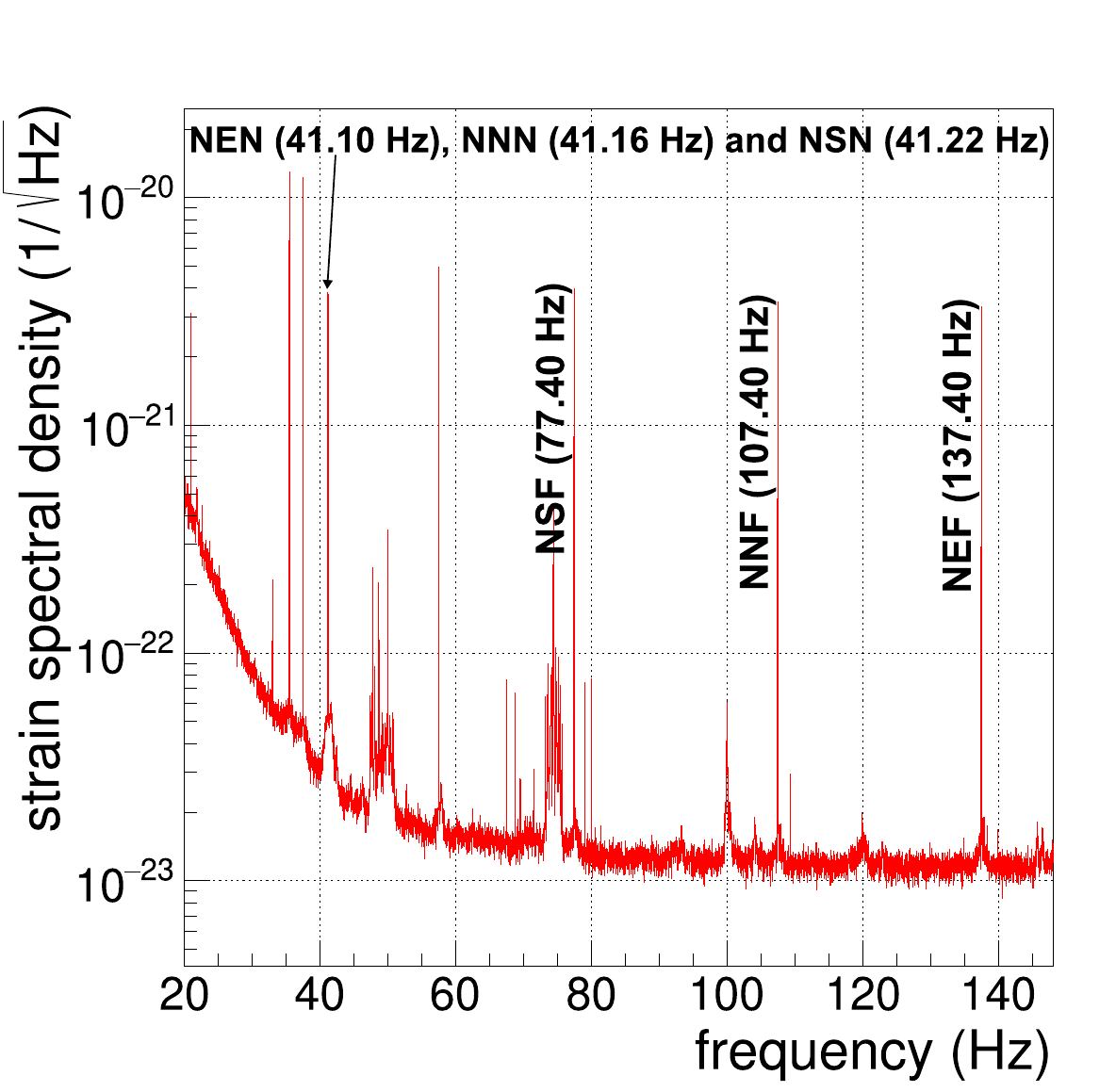}\label{fig:spectrum_a}}
  \hfill
  \subfloat[]{\includegraphics[width=0.5\textwidth]{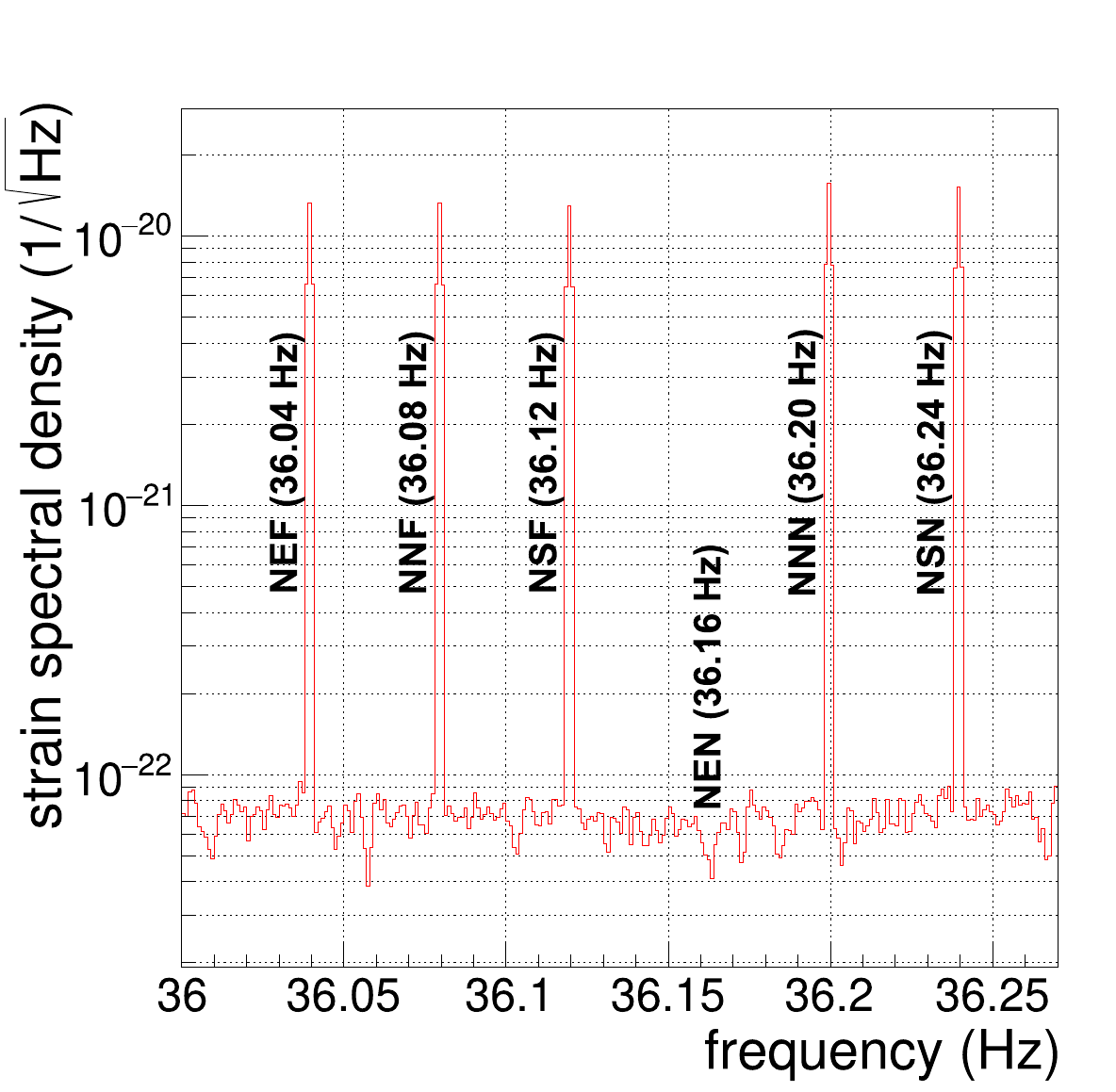}\label{fig:spectrum_b}}
    \caption{Example of a frequency spectra of Virgo interferometer strain data, on which multiple calibration lines have been injected using the six NCals. 
    NEN was set at an angle $\psi = 90^{\circ}$.
    Figure (a) data was taken on March 23, 2024, between 16:09 and 16:54 UTC. 
    The frequency resolution is $10~\mathrm{mHz}$.
    Figure (b) shows a $1~\mathrm{mHz}$ resolution spectrum in the nominal operating bandwidth of the NCal system Data was taken on March 30, 2024, between 06:56 and 08:36 UTC.}
    \label{fig:spectrum}
\end{figure}

An isometric view of a rotor is shown in \autoref{fig:rotor}.
It is divided into two $90$-degree sectors.
Each sector is a quarter of a cylinder with a nominal diameter of $208~\mathrm{mm}$ and thickness of $104.4~\mathrm{mm}$. 
The mass of a rotor is $5~\mathrm{kg}$ for those made of aluminium and $2.5~\mathrm{kg}$ for those made of PVC.
Such two sectors rotor induces a signal at twice the rotation frequency in the interferometer strain $h(t)$.
\autoref{fig:spectrum} shows typical Virgo sensitivity spectra, a few weeks before the start of the O4b run (Virgo was still doing commissioning during the first part of the O4 run). 
Unlike during standard operation, the NCal signals (or calibration \textit{lines}) were not removed on the reconstructed $h(t)$ data used for this figure in order to make them well visible.
The zoomed spectrum in figure \ref{fig:spectrum_b} illustrates the stability of the injected lines.
With the chosen frequency resolution, signals injected by NCals exceed the detector nominal sensitivity by few orders of magnitude in the low-frequency band.

For practical and safety considerations, rotors are encapsulated in aluminium housings.
Ball bearings and a motor are installed on the rotor axis. 
The frequency and the phase of the rotor are retrieved using a photodiode, capturing signal emitted by a LED located on the other side of the rotor.
The luminosity measured by the photodiode decreases significantly as the rotor passes in front of the LED.
To distinguish sectors, one side of the rotors has been sandblasted for aluminium rotors or painted white for PVC rotors, which changes its reflectivity.
The rotor frequency and phase are measured with this system every half turn.
Feedback loops have been developed, as part of the Virgo real time control system, to synchronize the frequency and phase of all rotors to the GPS timing system, with a short-term phase jitter between $1$ and $2~ \mathrm{mrad~RMS}$, but much more stable once averaging over a few turns.
\\

Both Near and Far rotors on the same side of the mirror are fixed on the same support.
These supports are suspended from a scaffolding not connecting to the vacuum chamber, in order to limit the propagation of parasitic rotor vibrations to the mirror environment.
The positions of these supports are continuously monitored to detect fine layout changes or possible hardware failures.\\

\subsection{Choice of orientation}

\begin{figure}[h]
    \centerline{\includegraphics[width=0.8\textwidth]{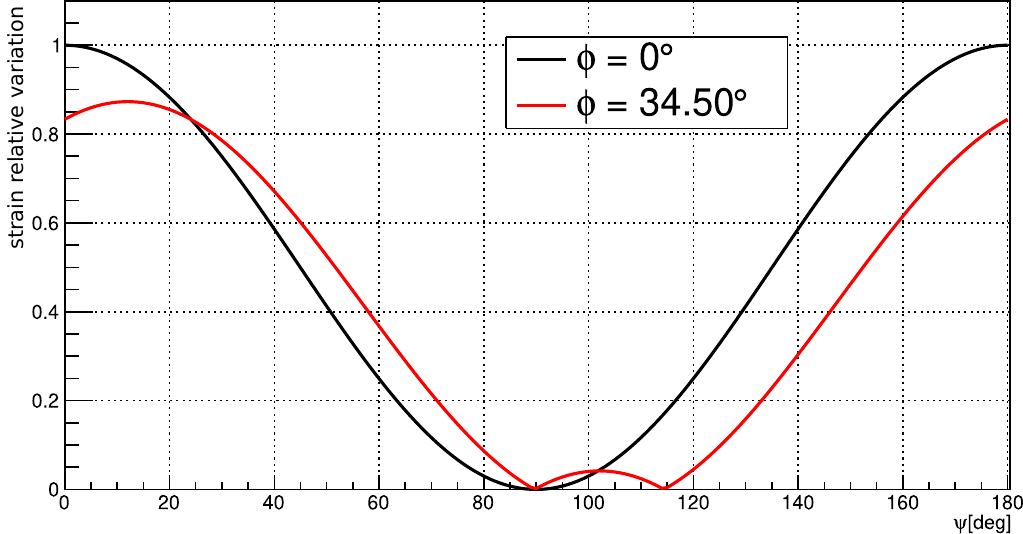}}
    \caption{Relative amplitude injected by a nominal Near NCal (at $1.7~\mathrm{m}$ from the mirror), for two orientations $\phi = 0^{\circ}$ (in black) and $34.5^{\circ}$ (in red) as function of the twist $\psi$, relative to maximal amplitude at $\phi = \psi = 0^{\circ}$}
    \label{fig:rotorTiltAndPhi_1D}
\end{figure}

The simulation tool FROMAGE \cite{NCalO3} has been used to compute the signal induced by a NCal at different angles $\phi$, between the line connecting the NCal support to the mirror and the interferometer beam axis, and twists $\psi$ of the rotor on the rotor-mirror axis (see figure \ref{fig:O4setup} for angles definition).
\autoref{fig:rotorTiltAndPhi_1D} shows the relative amplitude of the signal injected by a NCal as function of the twist $\psi$, for an unrealistic value of $\phi = 0^{\circ}$ and a more suitable one of $\phi = 34.5^{\circ}$.
Nominal values of $\phi = 34.5^{\circ}$ and $\psi = 12^{\circ}$ have been selected for all rotors as they maximize the signal and minimize its uncertainty.

\subsection{NCal operation}

The configuration presented in this paper yields redundancies and robustness. 
If a NCal stops working, we will still be able to perform the calibration with the remaining actuators.
The ability of such a NCal system to operate continuously for several months with minimum human intervention was tested prior to the run. 
During O4, it is used to continuously inject calibration signals at constant frequencies of about $36~\mathrm{Hz}$.
To identify each NCal signal, their frequencies are slightly shifted by a few hundredths of a hertz.
To minimize the science data contamination, NCal lines are then subtracted from the reconstructed strain data $h(t)$ \cite{Virgo_Cali}, using their predicted signals time series.

\section{Positioning uncertainties} \label{sec:setupUncertainties}
This section describes the uncertainties related to NCals positioning. 
The resulting values will be combined in table \ref{table:uncertainties_summary} with the other sources of errors to provide the overall NCal system uncertainty.
A precise knowledge of all NCals position is mandatory to make accurate predictions.
Results presented in this paper are based on two geometrical surveys carried out two years apart, in 2021 and 2023.
On these occasions, the positions of the NCal reference plates were measured relative to the building reference frame around the mirror vacuum chamber.
By combining the measurements from these two studies and transferring the reference plates position to the suspended NCals, we have been able to locate the NCals precisely and measure the associated uncertainties \cite{geoSurvey}.

\subsection{NCal to mirror distance} \label{subsec:NCalToMirror}

To determine the position of the north-end mirror relative to the NCal system, we injected signals with North, South and East rotors, at slightly shifted frequencies. 
The mirror position can be computed by minimizing the difference between the observed and predicted (using FROMAGE simulation) NCal lines.
Applying this procedure on Far NCals, we were able to measure the bias between the position of the mirror and the center of the NCal system to be $-5.48~\mathrm{mm}$ on the x-axis and $-1.95~\mathrm{mm}$ on the y-axis, with an uncertainty of $1.2 ~\mathrm{mm}$.
This uncertainty value is dominated by systematics on the injected NCal signals, of typically $0.2\%$ (see section \ref{sec:allUncertainties}).

\begin{figure}[h]
  \centering
  \subfloat[]{\includegraphics[width=0.5\textwidth]{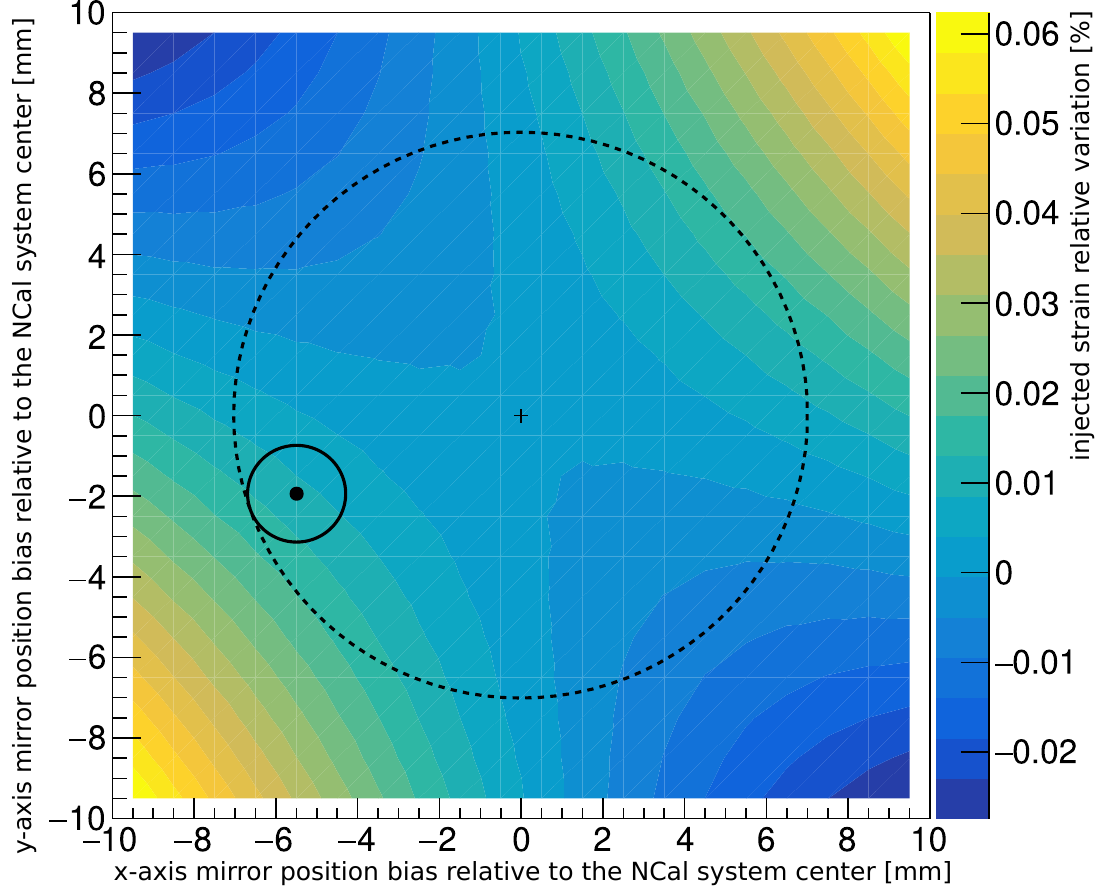}\label{fig:1.7m_NS_strain}}
  \hfill
  \subfloat[]{\includegraphics[width=0.44\textwidth]{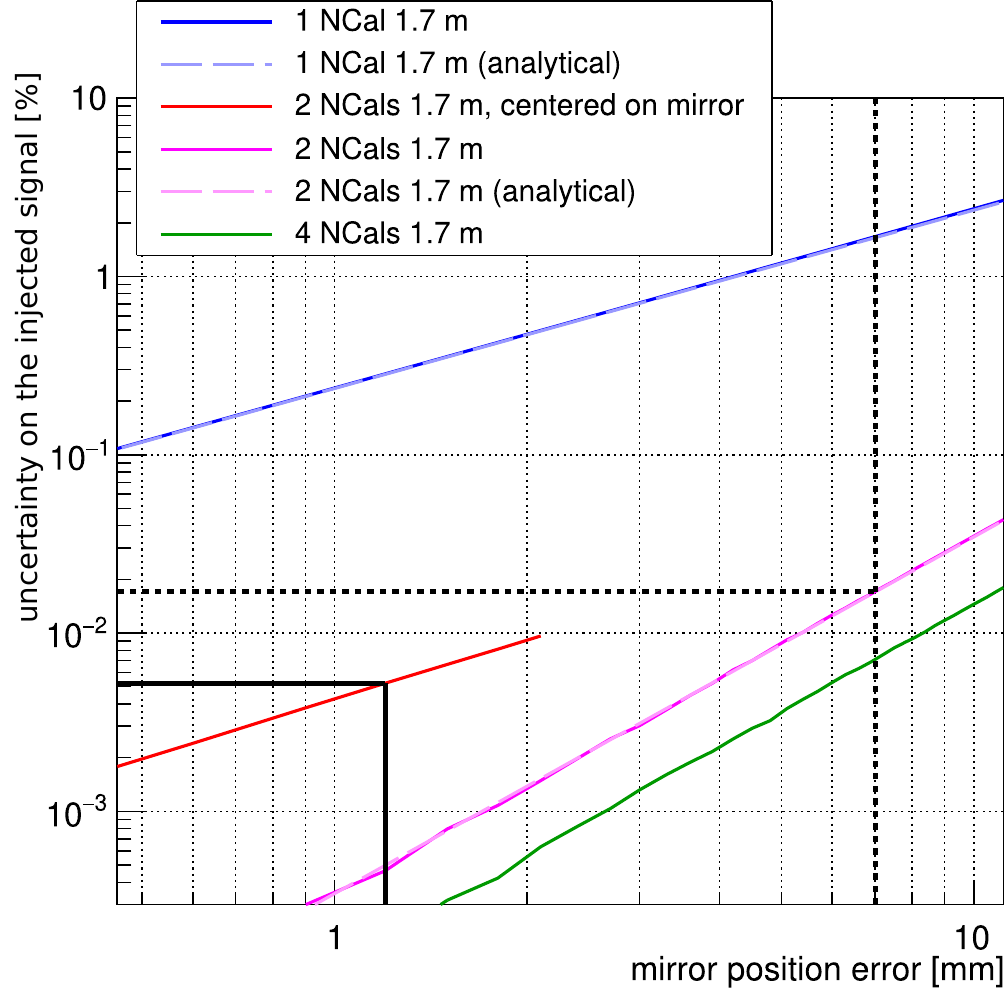}\label{fig:1.7m_NS_radii}}
  \caption{
  (a) Relative variations of the averaged Near South and North NCal amplitudes as function of the mirror position. 
  The coordinates are such that the North NCal is in the lower left corner and the South NCal in the upper right corner.
  The solid black circle is centered on the mirror location measured using Far NCals, and represents the associated uncertainties.
  The dashed black circle gives the uncertainties without using the measured mirror offsets.
  (b) Maximum error in the injected amplitude as function of the error in mirror position, for one, two and four NCals setups. See text for details.
  }
  \label{fig:1.7m_NS}
\end{figure}

\autoref{fig:1.7m_NS_strain} shows the relative variation of the averaged amplitudes injected by two NCals, computed by the FROMAGE simulation, as function of the mirror position.
The black dot represents the computed position of the mirror, with the associated uncertainty represented by the solid black circle. The dashed circle represents a conservative uncertainty that uses the computed position without correcting for it.
\autoref{fig:1.7m_NS_radii} shows the maximum signal variation as function of the uncertainty on the mirror position.
In this figure, all curves except the red one assumed the mirror to be at the mechanical center of the system (black cross of figure \ref{fig:1.7m_NS_strain}). 
Solid curves are produced using the FROMAGE finite element simulation, while the dashed ones represent analytical predictions made by considering point-like mirror and twistless rotors.
On this figure, analytical curves are difficult to see because they are just behind the numerical simulation.
Black vertical lines correspond to the radii of circles in figure \ref{fig:1.7m_NS_strain}.
Without correction for the mirror position offset, the intersection between the purple curve and the black dashed line gives an uncertainty of $0.02\%$ on the injected signal.
This value is two orders of magnitude lower than that obtained with a single NCal system (in blue), which could not measure the mirror position.
By correcting the position of the mirror with the offset we measured, we managed to reduce this uncertainty to $0.005\%$, indicated by the intersection of the red curve and the continuous black line in figure \ref{fig:1.7m_NS_radii}.
Although this is not needed given the other source of uncertainties, the error coming from the mirror position could be further reduced by another order of magnitude by moving the NCal setup by few millimeters to center it on the measured mirror position.
Let us remark that once each NCal has been cross-calibrated with respect to the two-NCal system, their systematic uncertainty coming from the mirror position is the one of the two-NCal system. 
This is providing a reliable accurate system, even if only one NCal is used after cross-calibration.

Recent work \cite{NCal_LIGO_2023} has suggested that using four actuators at the four cardinal points and averaging their signals could be a way to mitigate this uncertainty.
Comparing the purple and the green curves of \autoref{fig:1.7m_NS_radii}, this approach decreases the uncertainty associated with the mirror's position by only a factor of $2$. 
On the other hand, it is more effective to measure and correct for the mirror position, even with just two NCals.

\vspace{-15pt}

\subsection{NCal to NCal distance} \label{subsec:NCalToNCal}

The previous subsection discussed the uncertainty coming from the mirror position within a perfectly known NCal system. 
However, when doing the geometrical survey and transferring measurements from different reference points, we measured the distance between each NCal with an accuracy of $0.5~\mathrm{mm}$.
The distance variation due to building temperature changes, less than $70~\mu m$, has a negligible contribution.
Using a first-order expansion of the distance $D$ dependency to the fourth power of the gravitational signal emitted by a NCal \cite{NCalO3}, this value can be converted into an uncertainty of $4\delta D / D = 0.12\%$, and $0.1\%$, on the injected signal by the Near and Far setups, respectively.
These results have been confirmed by the FROMAGE simulation.
\\

\subsection{NCal supports to beam axis angle ($\phi$) and NCal twist ($\psi$)} \label{subsec:PhiAndPsi}

Geometrical surveys were also used to determine the values of the angles $\phi$ and $\psi$, defined in section \ref{sec:O4system}, along with their associated errors.
When predicting the NCal signal amplitudes, we used these measured values of $\phi$, with uncertainties of $0.06^{\circ}$.
By deriving the $\cos(\phi)$ dependency of the force induced by the NCal \cite{NCalO3} for small $\delta \phi$ angles, this translates into a relative uncertainty of $\delta \phi \tan(\phi) = 0.07\%$ for each NCal.
Therefore the uncertainty on the signal injected by a two NCal setup is $0.07 / \sqrt{2} = 0.05\%$.

Using the measured $\phi$ angles, we recalculated the optimum twists $\psi$.
These values are found to be close to $12^{\circ}$, which is why this value was kept.
By adding the possible machining asymmetry of the supports to their measured misalignment relative to support-mirror axis, we constrained the error on twists to be within $0.08^{\circ}$.
Thanks to the choice of the nominal value of the twist (see \autoref{sec:O4system}), the impact of such an error on the injected signal is of the order of $0.001\%$.

Additional errors in $\phi$ and $\psi$ due to incorrect positioning of the mirrors are already taken into account in the subsection \ref{subsec:NCalToMirror} when moving the mirror position.

\subsection{NCal vertical position} \label{subsec:NCalVertical}

The elevation of the NCals relative to the interferometer plane leads to a deviation in the injected signal.
We estimate the vertical offset of the NCals to be less than $z_{\max} = 10~\mathrm{mm}$.
The amplitude of the signal induced by a rotor varies with the fourth power of the NCal to mirror distance ($d \sim D/2$).
Therefore, a small vertical offset reduces this force by $2 (z_{\max}/d)^2$, which, once projected in the interferometer plan, constrains the signal deviation to be less than $5/2\left(z_{\max}/d \right)^2 = 0.008\%$ for the Near setup and $0.005\%$ for the Far setup.

These uncertainties can be further reduced by better estimating the NCal vertical offset.
Previous work has shown that this offset can be derived from measuring the phase difference with two NCal rotating in opposite direction \cite{NCalO3}, or by "twisting" one NCal by $180^{\circ}$.
However, as this parameter is not dominant, we have decided to keep the above values.

\section{Rotor uncertainties} \label{sec:rotorUncertainties}
This section describes the uncertainties coming from rotor knowledge. 
As for the position uncertainties, the resulting values will be combined in table \ref{table:uncertainties_summary} with the other sources of errors to provide the overall NCal uncertainty.

The FROMAGE simulation is used to calculate the gravitational force applied by a rotor to the mirror.
It discretizes the rotor and the mirror into 3D finite elements, computes and integrates the gravitational force induced by each rotor element on each mirror element.
The systematic uncertainties due to this modelling method are described in this section.
Detailed studies on the systematics are publicly available in Virgo technical notes for the NNN \cite{NNN-R4-10}, NNF \cite{NNF-R4-02}, NSN \cite{NSN-R4-12}, NSF \cite{NSF-R4-03}, NEN \cite{NEN-R4-11} and NEF \cite{NEF-R4-06} rotors.

\subsection{Grid uncertainty}
We set the resolution of the FROMAGE simulation grid in order to limit the numerical error to $0.005\%$ for Near NCals. 
The number of mirror and rotor elements are respectively $2880$ and $41600$.
This error is an upper limit for Far NCals, for which a larger mirror to rotor distance decreases the uncertainty from the discretization.

\subsection{Rotor geometry uncertainty}
In order to model each rotor in the software, we made multiple measurements.
The sectors of each rotor are divided into $16$ sub-sectors.
Accounting for systematics due to the measurement technique and statistical variations, sub-sector thicknesses are known to within $1.8 \times 10^{-2}~\mathrm{mm}$, and their radii within $1.4 \times 10^{-2}~\mathrm{mm}$.

We performed a Monte Carlo simulation by computing the gravitational signal emitted by a nominal rotor, for which the thickness and radius of each sub-sector vary randomly within the previous uncertainties. 
The resulting uncertainty on the mirror displacement due to rotor geometry is $0.025\%$.

\subsection{Sector asymmetry uncertainty}
Thanks to the $90^{\circ}$ opening angle of each sector, the machining precision of these opening angles is not critical. Using FROMAGE simulation, the measured defects translate to signal a uncertainty of less than $0.001\%$.

\subsection{Global uncertainties} 
The rotors are machined from cylindrical blocks, made from either PVC or aluminium.
Measurements of the mass and dimensions of these cylinders have allowed us to constrain the density of these materials to within $0.2~\mathrm{kg.m^{-3}}$.
This corresponds to an uncertainty of $0.014\%$ on the PVC NCals (Near) signal and $0.007\%$ on the Al NCals (Far) signal.

Temperature variations, due to the heat dissipated by the NCal motors or fluctuations of the building air conditioning system, change the size of the rotors, which modifies the injected signal.
The monitoring of the NCals temperature over several months of continuous operation shows a stability to within $1.5~\mathrm{K}$. 
Given the coefficient of thermal expansion of the rotor materials, the resulting uncertainty on the signal is $0.024\%$ for PVC NCals and $0.007\%$ for Al NCals.

The uncertainty on the Newtonian constant of gravitation, taken as $G = 6.67430 \times 10^{-11} \pm 1.5 \times 10^{-15} ~\mathrm{m^{3}.kg^{-1}.s^{-2}}$ \cite{CODA_G}, induces an uncertainty of $0.002\%$ on the NCal signal. 

\subsection{Total rotor modeling uncertainties}
Overall, we obtain the systematic uncertainty resulting from the rotor modelling by calculating the quadratic sum of all the terms mentioned above, resulting in a value of $0.038\%$ for Near rotors and $0.027\%$ for Far rotors.
This is one order of magnitude better than O3 results \cite{NCalO3}, thanks to detailed rotor metrology and density measurements.
As the rotor geometries may be correlated due to the use of the same machining technique, a conservative estimate of the uncertainty of a two-rotor configuration made of the same material is to use the same relative uncertainty as for a single rotor configuration.

\subsection{Rotor balancing}
NCal rotation must be stable over long periods, with minimum vibrations.
Small asymmetries between the two sections of a rotor can cause a recoil motion of the rotor enclosure and support, and therefore produce a parasitic gravitational signal (see \autoref{sec:couplingUncertainties}).
For this reason, we balanced each rotor with small aluminium or PVC plates acting as a counterweight.

\subsection{Elastic deformation}
\begin{figure}[h]
    \centerline{\includegraphics[width=0.9\textwidth]{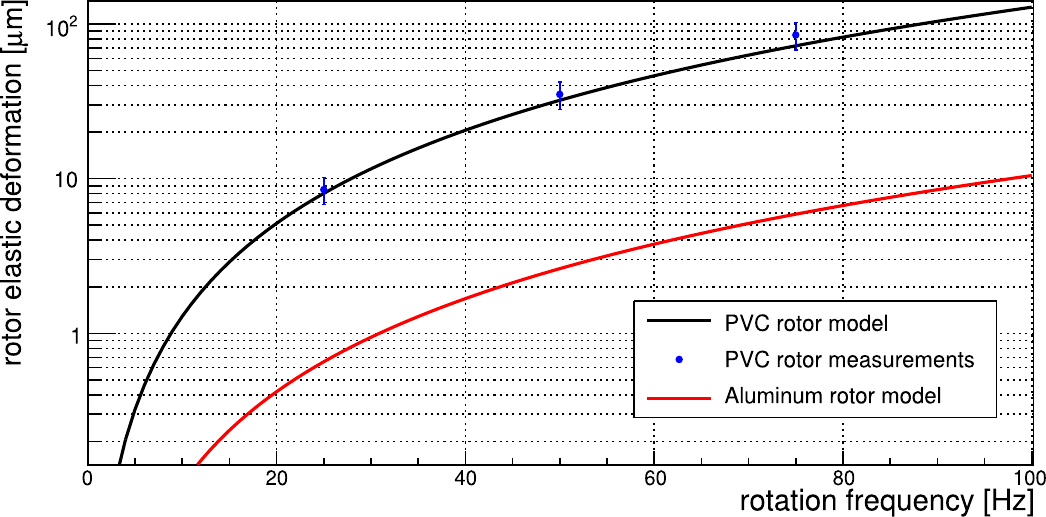}}
    \caption{Elastic deformation of a section of PVC and aluminum rotors due to centrifugal force as a function of rotation frequency.}
    \label{fig:elastic_elongation}
\end{figure}

As the rotors spin, a centrifugal force tends to deform them.
This has the effect of changing their geometry, and therefore modifying the induced gravitational signal.
A finite element simulation, along with a simple analytical model, predicts at $18~\mathrm{Hz}$ an elastic elongation of about $0.4~\mathrm{\mu m}$ for aluminium rotors and $5.6~\mathrm{\mu m}$ for PVC rotors.
As shown in figure \ref{fig:elastic_elongation}, for PVC rotors we managed to confirm experimentally this model to within $20\%$, by measuring the spacing between the rotor and its enclosure with a LED/photodiode system.
By testing several frequencies, we were also able to check that elongation does vary as the square of the rotation frequency.
Simulations of rotors with such deformations indicate that the NCal signal is modified by $0.022\%$ for PVC rotors, and by less than $0.01\%$ for aluminium rotors, at $18~\mathrm{Hz}$.

Although the rotor expansion is a known and predictable effect, given the fairly low value at our usual operating frequency, we are not correcting for it, and include this effect as uncertainty.
As a side remark, at too high rotation frequency, a permanent plastic deformation of the material is expected. This is the reason why we limit the rotation speed of PVC rotor to $50~\mathrm{Hz}$ and only use Aluminium rotors for higher frequencies.

\section{Residual parasitic coupling uncertainties} \label{sec:couplingUncertainties}
In addition to the desirable gravitational effect, NCals can couple to the mirror in various ways.
In this section, we discuss the effects of these parasitic couplings, their estimation and mitigation actions.
The study of the magnetic field produced by the rotors turned out to be a key part of this study.
Again, the resulting values will be combined in table \ref{table:uncertainties_summary} with the other sources of errors to provide the overall NCal uncertainty. 

\subsection{Magnetic coupling}

Small magnets are attached to the mirror to control their position, making them sensitive to parasitic magnetic fields. 
For this reason, the magnetic field in the Virgo buildings is monitored by environmental sensors.
In addition, we have installed dedicated coils around Virgo's north-end mirror to monitor the fields produced by the NCals and inject test signals.
All these coils lie in the same plane as the interferometer.

These sensors observed a tiny magnetic field of a few tenths of $\mathrm{nT}$ at twice the NCal rotation frequency.
This induced a small motion of the mirror and should therefore be considered as parasitic coupling.
We used small coils installed on some NCal housing to simulate the effect on the mirror of NCal magnetic lines with the NCal at rest.
The resulting mirror displacement was measured and compared with the predicted NCal gravitational effect.
The effect of the parasitic magnetic coupling relative to the signals injected by a Near NCal has been measured to be up to $0.5\%$.
\\

The original O4 NCal setup was composed only of aluminium rotors.
However, as this material is known to be conductive, under the effect of the ambient magnetic field, eddy currents do form in the moving rotor. 
These currents then produce a variable field, opposite to that of the environment, which can explain the observed magnetic coupling.
If this is the case, using a different material can reduce magnetic coupling.
Therefore, we decided to replace some Al rotors with PVC rotors.
This reduced the magnetic field by two orders of magnitude compared to aluminium rotors making it negligible.
As PVC NCals are half as massive as aluminium ones, the induced gravitational amplitude is also reduced by a factor of $2$.
Therefore we installed them on the Near locations where the injected signal is about two times larger than for the Far location. 
One down side of PVC is its elasticity, already mentioned in \autoref{sec:rotorUncertainties}. 
Rotation speeds larger than 50 Hz can lead to permanent deformation of the rotors.
For these reasons, we kept aluminium rotors at the Far locations, to be able to inject gravitational signals up to $150~\mathrm{Hz}$ for the start of O4b. 
We also kept on investigating how to reduce their magnetic field.

A simple solution deployed during the NCal commissioning was to wrap them in magnetic field shielding film (MCL61 from YSHILD®), reducing the magnetic field by a factor $3$.
The uncertainty due to magnetic coupling on the calibration signal injected by the shielded Far NCals has then been measured to be at most $0.2\%$ \cite{Note_Mag}.
In the future, other magnetic shielding solutions may be considered, such as the replacement of aluminium housing by iron boxes.

Magnetic field compensation has also been explored.
Using two coils centered on a NCal, it is possible to generate a magnetic field that counteracts the one generated by the rotor.
Although this approach seems attractive, it requires precise knowledge of the field at the exact position of the mirror in order to cancel it out effectively, something that is currently not possible.


\subsection{Gravitational coupling with the NCal supports}

The NCals are mounted on aluminium supports, suspended from the Virgo infrastructure to damp NCal induced vibrations.
Position sensors track support motions, measuring amplitude up to half a micrometer at twice the rotation frequency.
These vibrations, caused by residual rotor imbalance, can then induce a gravitational coupling between the entire setup and the mirror, adding an uncertainty on the calibrator effect.
\\

However, what we measure is the recoil motion of the NCal support due to small unbalance defaults of the rotor. 
Therefore the suspended center of mass is not moving. 
But since the masses distribution of the rotor and its axis, enclosure and suspending frame are not the same, we expect some effect on the mirror. 
Given the observed amplitude, a first order model of the relative masses distribution predicts a coupling of the order of $10^{-4}\%$ of the direct NCal signal at twice the rotor frequency.
This model has been checked at the rotation frequency, where the NCal signal emitted by the rotors is negligible, and was not seen in the interferometer ouput.

\subsection{Coupling with the mirror suspension}

The last stage of the mirror suspension, the so-called \textit{marionette}, is also sensitive to the NCals.
Since it is further away than the mirror, we can use the mirror motion as an upper limit of the marionette motion.
This motion is filtered by marionette-to-mirror transfer function, which can be modelled by a simple pendulum with a resonance frequency of $f_0 = 0.6~\mathrm{Hz}$.
At $36~\mathrm{Hz}$, this results in a parasitic motion of $0.03\%$ of the direct mirror’s motion induced by the NCals. 
It is included as part of the residual coupling uncertainties.

\subsection{Coupling with the induced torque}

Due to the NCal positioning relative to the laser beam axis, all the mirror parts are not subjected to the same force.
Therefore an NCal induces a torque on the mirror which rotates around its center of mass.
As the interferometer beam is not perfectly centered on the mirror, this torque results in an optical path difference and bias the calibration signal.
Assuming a typical offset of $0.5~\mathrm{mm}$ between the beam and the mirror center, the torque produced by a Near (respectively Far) NCal results in signal variation of $0.03\%$ ($0.025\%$).
However, this torque either increases or decreases the injected signal, depending whether the NCal is at the front or back of the mirror. 
Therefore, for a two NCal system, the torque effect cancels out, and the system is not sensitive to beam misalignment.
If the two NCal system is not perfectly centered on the mirror, the torques induced by each NCal are not anymore identical,leading to a residual effect.
With our measured offset, the signal variation, considered as an uncertainty, of about $0.003\%$.

\subsection{Residual parasitic coupling}

\begin{figure}[h]
    \centerline{\includegraphics[width=1\textwidth]{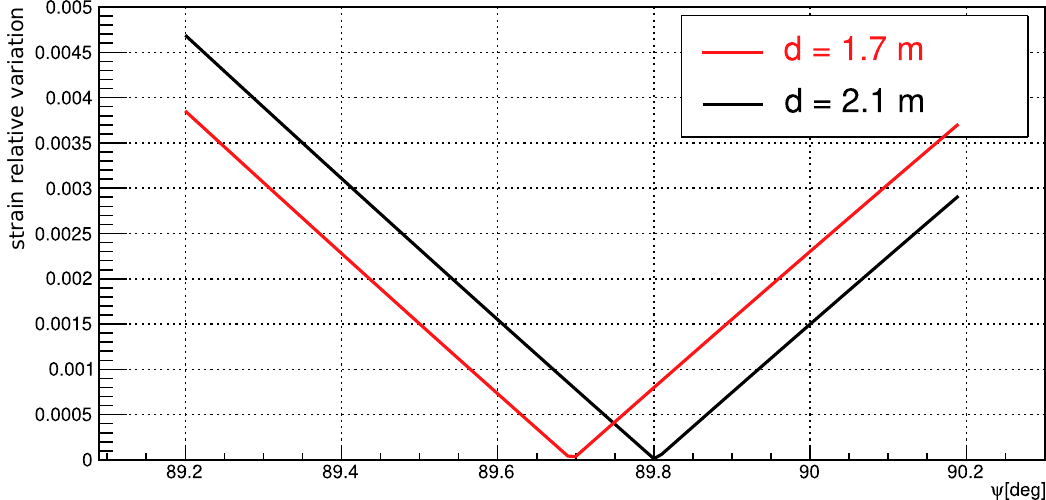}}
    \caption{Relative amplitude injected by NCals at $1.7~\mathrm{m}$ (in red) and $2.1~\mathrm{m}$ (in black) from the mirror, for an orientation $\phi = 34.5^{\circ}$, as function of the twist angle $\psi$, relative to maximal amplitude at $\psi = 12^{\circ}$.}
    \label{fig:90twist_34.5}
\end{figure}

Possible other couplings can induce an unpredicted mirror motion.
\autoref{fig:rotorTiltAndPhi_1D} and \autoref{fig:90twist_34.5} show that there are values of twist $\psi$ for which the induced gravitational signal becomes null. 
Therefore, at these angles, residual parasitic couplings to the mirror can be probed.
For this purpose, we have set the twist of Near East NCal to $89.7^{\circ} \pm 0.1^{\circ}$.
The remaining gravitational effect of the NCal is expected to be less than $0.1\%$ of the nominal signal produced by a similar rotor with a $12^{\circ}$ twist.
In addition, for this value of twist, measurements in laboratories predict that magnetic coupling is inferior at $0.005\%$, and parasitic signal due to displacement of the setup is expected to be below $0.001\%$.

At this twist value, we experimentally observed a signal of $0.1\%$ of the nominal NCal signal in the spectrum of the interferometer. 
Given that this residual signal aligns with the expected uncertainty arising from the twist, further limiting the parasitic coupling for Near NCals may be challenging. 
In the case of Far NCals, magnetic coupling dominates over the other residual effects, raising this value to $0.2\%$, values which therefore are reported in table \ref{table:uncertainties_summary}.

\section{Overall uncertainty budget and experimental validation} \label{sec:allUncertainties}
This paper presents a detailed study of the NCal calibration uncertainties.
\autoref{table:uncertainties_summary} summarises those uncertainties considered for the O4 NCal setup and presented in the previous sections.
As each term is assumed to be independent, the overall uncertainty is taken as their quadratic sum.
\\

\begin{figure}[h]
    \centerline{\includegraphics[width=0.7\textwidth]{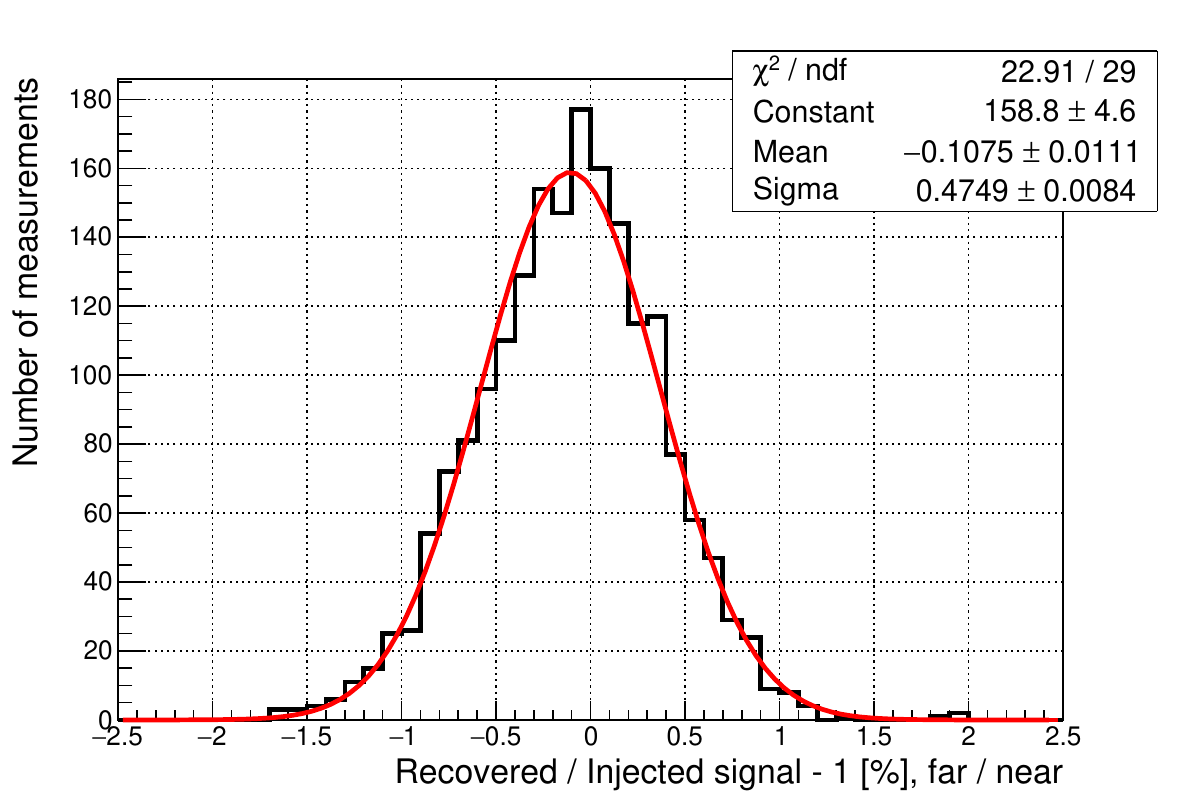}}
    \caption{Comparison of the calibration ratios between the Near and Far NCal setups using data from March 15, 2024. 
    Measured relative differences, in black, correspond to a fitted normal distribution, in red.
    Calibration lines where spanning from $40.98$ to $41.12~\mathrm{Hz}$ for this measurement. 
    There is one data point every $100$ seconds.}
    \label{fig:hrec_hinj_near_far}
\end{figure}

During commissioning and later during the run, one of the key calibration results is the ratio between the recovered and the injected calibration line amplitude ($h_\mathrm{rec}/h_\mathrm{inj}$).
To minimize the effect of the NCal to mirror distance uncertainty, we usually use the average value of this ratio for the North and South NCals. 
Since the Near and Far NCals are made of different material and have systematic uncertainties only partially uncorrelated, especially for the parasitic coupling, comparing their results is one of the key consistency checks of the NCal calibration.
\autoref{fig:hrec_hinj_near_far} shows this comparison on a day of data from the 16th engineering run, March 15, 2024, prior to the start of the O4b run.
The offset of the ratio between the Near and Far NCals is around $0.11\% \pm 0.01\%$, which is consistent with the values given in \autoref{table:uncertainties_summary}.

\begin{table}[!h]
    \begin{center}
        \resizebox{\columnwidth}{!}{
            \begin{tabular}{||c|c|c|c|c||} 
                \hline\hline\hline
                \multicolumn{2}{||c|}{Parameter} & Formula & $h_\mathrm{rec} / h_\mathrm{inj}$ Near [\%] & $h_\mathrm{rec} / h_\mathrm{inj}$ Far [\%]\\
                \hline\hline\hline
                \multirow{5}{*}{Positioning} 
                 & NCal to NCal distance ($D$) & $4 \delta D / D$ & $0.12$ & $0.10$\\
                 & NCal supports to beam axis angle ($\phi$) & $\delta \phi \tan(\phi) / \sqrt{2}$& $0.05$ & $0.05$\\
                 & NCal to mirror distance ($d \sim D/2$) & see \autoref{sec:setupUncertainties} & $0.005$ & $0.005$\\
                 & NCal twist ($\psi$) & see \autoref{sec:setupUncertainties}  & $\leq 0.001$ & $\leq 0.003$\\
                 & NCal vertical position ($z$) & $5/2 (z/d)^2$ & $0.008$ & $0.005$\\
                \hline
                \multicolumn{2}{||c|}{
                Rotor modeling
                } & see \autoref{sec:rotorUncertainties} & $0.038$ & $0.027$\\
                \hline
                \multicolumn{2}{||c|}{Rotor elastic deformation at $18~\mathrm{Hz}$ ($36~\mathrm{Hz}$ in $h(t)$)} & see \autoref{sec:rotorUncertainties} & $0.02$ & $\leq 0.01$\\
                \hline
                \multicolumn{2}{||c|}{Residual coupling (including magnetic)} & see \autoref{sec:couplingUncertainties} & $\leq 0.1$ & $\leq$ $0.2$\\
                \hline\hline\hline
                \multicolumn{2}{||c|}{Total} & quadratic sum & $0.17$ & $0.23$\\
                \hline\hline\hline
            \end{tabular}
        }
    \end{center}
    \caption{
        $1\mbox{-}\sigma$ uncertainty budget, in percent, on calibration signal amplitude for the Near and Far North-South NCal pairs used at the start of O4.
        The positioning uncertainties, discussed in \autoref{sec:setupUncertainties}, cover the position of the mirror relative to the NCal, the distance between the North and South NCals and the rotor orientation.
        The rotor induced strain uncertainty, discussed in \autoref{sec:rotorUncertainties}, encapsulates all the uncertainties involved in FROMAGE's prediction of the signal produced by rotors.
        The rotor deformation under the effect of centrifugal force due to rotation is also discussed in \autoref{sec:rotorUncertainties}.
        The residual couplings are discussed in \autoref{sec:couplingUncertainties}.
        The total uncertainty is defined as the quadratic sum of all the contributions. 
    }
    \label{table:uncertainties_summary}
\end{table}

\section{Conclusion and outlook} \label{sec:conclusions}
In this paper we describe the NCal installation used on the Virgo detector at the start of the O4b run.
This system of multiple rotors made of different material provided redundancy. 
Comparisons prior to the start of the run between the Near and Far NCals gave a result well within the systematic uncertainties.
These uncertainties on the injected mirror displacement are as low as $0.17\%$ ($1\mbox{-}\sigma$ confidence interval) for the Near NCal pair at the nominal frequency of the NCal injected signals (around $36~\mathrm{Hz}$).
This is a factor of $2$ below the PCal combined standard uncertainty \cite{LIGOPCal}.
Therefore, just before the start of the run, we used NCals to adjust the absolute calibration of PCals, which can explore higher frequencies.
We will keep monitoring the stability of the NCal system during the O4 run.

Although the NCal system was continuously running for many months prior to the start of the run, their operation during a year-long run will be a challenge and should bring valuable information for its improvement. 
We plan to replace the north and south Far aluminum rotors by PVC ones and possibly polyether ether ketone (PEEK) to mitigate the magnetic coupling and elastic deformation.
The uncertainty on the signal injected by the Far setup should then drop below $0.15\%$ bringing new interesting crosschecks with the Near NCals.
For the future, other solutions, such as shielding, are being considered to further reduce magnetic coupling, without affecting the system ability to inject signals at frequencies above $100~\mathrm{Hz}$.
With the current configuration, due to poor resolution in the twist angle, we are unable to constrain the residual coupling to a level below $0.1\%$.
We are therefore investigating options to improve this measurement.

As concluding remarks, the NCal system presented in this paper shows significant improvements compared to previous results obtained with prototypes. 
It is currently the system injecting a calibrated mirror motion on a GW interferometer with the lowest systematic uncertainties.
An advantage of the NCal system is that it relies primarily on mass and distance measurements, for which commercial measuring tools are more accurate than for power measurements required for calibrating PCals, avoiding the need for cross calibration with metrology institutes.
Further significant improvements are still possible and could be made in the coming years. 

\section*{Acknowledgments}
We are indebted to the Virgo Collaboration for allowing us to use the data collected during the tests and engineering run reported here. 
We are grateful for support provided by the Virgo Collaboration and the European Gravitational Observatory during those tests. 
We thank our colleagues in the Virgo Collaboration and in the LIGO Scientific Collaboration for useful discussions. 
We thank the technical staff at IPHC for their help in building the O4 NCal system.
Part of this work was supported by the grant ANR-21-CE31-0024 “ACALCO” from the French National Research Agency (ANR).

\end{document}